\documentstyle{article}

\begin{document}

\title{New transformation law for Heun and Hypergeometric Equations.}
\author{Yves Gaspar\\
DAMTP, Centre for Mathematical Sciences,\\
Cambridge University, Wilberforce Road, \\
Cambridge CB3 0WA, UK\\
\it E-mail: yfjmg@yahoo.co.uk}
\maketitle

\abstract{In this work we establish new forms of Heun-to-Heun transformations and Heun-to-Hypergeometric transformations. The transformations are realised by changing the independent variable in a non-linear way. Using these we also point out some simple examples of transformations between equations that are not Fuchsian and that generalise the Heun-to-Hypergeometric transformations.}

\section{Introduction.}

Heun's differential equation is a generalisation of the Hypergeometric 
equation : the latter has three regular singular points, while Heun's equation has one more finite regular singularity. The hypergeometric function is characterised by three parameters and an indepedent variable, while the solutions to Heun's equation depend on six parameters and an independent variable.
Any linear second order ordinary differential equation with four regular singular points can be reduced to Heun's equation.
This equation can arise in several problems of mathematical physics characterised by non-monotonous inhomogeneity profiles, such as diffusion, 
wave propagation, magneto-hydrodynamics and heat or mass transfer. A particular case of the Heun equation can be related to the Lam\'{e} equation. This equation plays an important role in problems of particle physics and cosmology of the very early universe \cite{Starob}, \cite{ivanov}. 

Recently results have been found \cite{maier} relating the Heun equation to 
the Hypergeometric equation. This reduction from the Heun to the Hypergeometric equation is realised by rational changes of the independent variable. These Heun-to-Hypergeometric transformations are analogous to the hypergeomeric identities as worked out by Goursat, following Riemann and Kummer.  

In this work we will establish some new Heun-to-Heun transformations. These transformations also include new Heun-to-Hypergeometric transformations. In all of the Heun-to-Hypergeometric transformations of \cite{maier} the independent variable is left unchanged, while in our work the {\it independent variable is changed in a non-linear way}. Most of the transformations for Heun and Hypergeometric functions studied in the literature are such that the independent variable is left unchanged, except for the cases where the independent variable is changed in a linear way.

We will also point out some simple transformations between equations having two regular singularities and one irregular singularity and equations having three regular singularities and one irregular singularity : this is in a sense a generalisation of the Heun-to-Hypergeometric transformations to the cases where the ordinary differential equations are {\it not Fuchsian}.           

\section{Some general identities for linear second order 
ordinary differential equations.}           

Given the functions $F(x)$ and $f(x)$, consider the following differential equations 

\begin{equation}
\label{aa}
{{{d^2}y}\over{dx^2}}-F(x){{dy}\over{dx}}-f(x)y=0
\end{equation}

\begin{equation}
\label{ab}
{{{d^2}u}\over{dx^2}}+[F(x)-{{df}\over{dx}}{{1}\over{f}}]{{du}\over{dx}}-f(x)u=0
\end{equation}

Any solution $y(x)$ to equation (\ref{aa}) can be used to find a solution $u(x)$ 
to equation (\ref{ab}) using the relation

\begin{equation}
\label{ac}
({{dy}\over{dx}}{{1}\over{y}})({{du}\over{dx}}{{1}\over{u}})=f(x)
\end{equation}

This can be shown easily by rewriting equation (\ref{aa}) by letting 
$$y(x)={e^{\int^{x}{h(t){dt}}}}$$

such that it becomes the Riccati equation

$${{dh}\over{dx}}+{{h(x)}^2}-F(x)h(x)-f(x)=0$$

Now if we put $h(x)={{f(x)}\over{v(x)}}$ in the above equation then we obtain 

$${{dv}\over{dx}}+{{v(x)}^2}+[F(x)-{{df}\over{dx}}{{1}\over{f}}]v(x)-f(x)=0$$

In this equation, if we substitute $v(x)={{du}\over{dx}}{{1}\over{u}}$ then we obtain equation (\ref{ab}) for $u(x)$, such that the relation  
$h(x)={{f(x)}\over{v(x)}}$ is equivalent to equation (\ref{ac}).

A similar calculation shows that the Riccati equation   

\begin{equation}
\label{aca}
{{dh}\over{dx}}-({{\alpha}\over{f}}){{h(x)}^2}+[F(x)-{{df}\over{dx}}{{1}\over{f}}]h(x)-f(x)=0
\end{equation}

and the second order equation 
\begin{equation}
\label{acb}
{{{d^2}y}\over{dx^2}}-F(x){{dy}\over{dx}}+{\alpha}y=0
\end{equation}
are related by the equation

\begin{equation}
\label{ad}
({{dy}\over{dx}}{{1}\over{y}})h(x)=f(x)
\end{equation}

The general identities (\ref{ac}) and (\ref{ad}) turn out to have interesting 
applications. A first application to the Hypergeometric equation will be discussed briefly in the next section.

\section{The Hypergeometric equation and a simple generalisation.}

Suppose that equation (\ref{aa}) is the hypergeometric equation, with
$$F(x)=-{{{c}\over{x}}-{{a+b-c+1}\over{x-1}}}$$
$$f(x)=-{{{ab}\over{x(x-1)}}}$$

Any second order Fuchsian ordinary differential equation with no more then three singular points 
can be reduced to the hypergeometric equation \cite{erdelyi}.
Let us work out equation (\ref{ab}) to obtain 

$${{{d^2}u}\over{dx^2}}+[{{x(1-a-b)+c-1}\over{x(x-1)}}]{{du}\over{dx}}+{{{ab}\over{x(x-1)}}}u=0$$
 
This is still a hypergeometric equation, but with $a \rightarrow -a$, $b \rightarrow -b$ and $c \rightarrow -c+1$ when compared with (\ref{aa}). \\

The non-linear transformation (\ref{ac}) thus relates two different hypergeometric equations.
Note that if $y(x)=F(a,b;c;x)$ is the solution to the hypergeometric equation (\ref{aa}) which is analytic at $x=0$ and equals unity at $x=0$, then it appears that $u(x)$ obtained from (\ref{ac}) will not be such a hypergeometric function with transformed parameters : it will not be analytic at $x=0$. This means $u(x)$ will represent a second linearly independent solution to the hypergeometric equation (\ref{ab}) ( see \cite{erdelyi} for a discussion about the two linearly independent solutions to the hypergeometric equation ). We will not discuss this in detail, note however that if $y(x)=F(a,b;c;x)$ then the first derivative of $y(x)$ is proportional to another hypergeometric function, see \cite{erdelyi},
$${{d{F(a,b;c;x)}}\over{dx}}={{ab}\over{c}}F(a+1,b+1;c+1;x)$$
such that 
$${{dy}\over{dx}}{{1}\over{y}}={{ab}\over{c}}{{F(a+1,b+1;c+1;x)}\over{F(a,b;c;x)}}$$

Such quotients of hypergeometric functions which then occur in (\ref{ac}), have been considered in the literature, see for instance \cite{R} and \cite{HH}.\\
\\

Now consider the case where the differential equation (\ref{aa}) is { \it not Fuchsian }, with   

$$F(x)=-{{{c}\over{x}}-{{a+b-c+1}\over{x-1}}}$$
$$f(x)=-{{{ab+mx}\over{x(x-1)}}}$$
This equation has two regular singularities located at $0$ and $1$ and an {\it irregular singularity} at $\infty$. Equations of this type are related to the Mathieu equation \cite{erdelyi}.  

Now for this case work out equation (\ref{ab})

$${{{d^2}u}\over{dx^2}}+[{{x(1-a-b)+c-1}\over{x(x-1)}}-{{m}\over{(ab+mx)}}]{{du}\over{dx}}+{{{ab+mx}\over{x(x-1)}}}u=0$$

This equation has {\it three} regular singularities located at $0$, $1$ and $-{{ab}\over{m}}$ and an irregular singularity at $\infty$. Thus in this case the non-linear transformation (\ref{ac}) relates two equations that are not Fuchsian and which have a different number of singularities. In the next section we study the effect of the non-linear transformation  (\ref{ac}) on the Heun equation.

\section{The Heun differential equation.}

The Heun differential equation is given by \cite{kamke}, \cite{snow}

\begin{equation}
\label{ae}
{{{d^2}y}\over{dx^2}}+({{\gamma}\over{x}}+{{\delta}\over{x-1}}+{{\epsilon}\over{x-d}}){{dy}\over{dx}}+{{{\alpha}{\beta}x-q}\over{x(x-1)(x-d)}}y=0
\end{equation}
where ${\alpha}$, ${\beta}$, ${\gamma}$, ${\delta}$, ${\epsilon}$, $d$ and $q$ are constants, and with the constraint ${\gamma}+{\delta}+{\epsilon}={\alpha}+{\beta}+1$. 
The four regular singular points are located at $x=0,1,d,\infty$.
The so-called 'trivial' Heun equations have ${\alpha}{\beta}=0$ and $q=0$. The parameter $q$ is called the accessory parameter.
First let us apply the transformation (\ref{ac}) to the above Heun equation and work out equation (\ref{ab}). One obtains 

$${{{d^2}u}\over{dx^2}}+[{{1-{\gamma}}\over{x}}+{{1-{\delta}}\over{x-1}}+ {{1-{\epsilon}}\over{x-d}}-{{{\alpha}{\beta}}\over{{{\alpha}{\beta}x-q}}}]{{du}\over{dx}}+{{{\alpha}{\beta}x-q}\over{x(x-1)(x-d)}}u=0$$

This equation has {\it five} regular singularities located at $x=0,1,d,{{q}\over{{\alpha}{\beta}}}$ and $x=\infty$. 

Note that if ${{q}\over{{\alpha}{\beta}}}=1$ then the above equation is another Heun equation with 
${\gamma} \rightarrow 1-{\gamma}$, ${\delta} \rightarrow -{\delta}$ and ${\epsilon} \rightarrow 1-{\epsilon}$.
If ${{q}\over{{\alpha}{\beta}}}=d$ then the above equation is a Heun equation with ${\gamma} \rightarrow 1-{\gamma}$,
${\delta} \rightarrow 1-{\delta}$ and ${\epsilon} \rightarrow -{\epsilon}$. If $q=0$ then the above equation is a Heun equation with ${\gamma} \rightarrow -{\gamma}$, ${\delta} \rightarrow 1-{\delta}$ and ${\epsilon} \rightarrow 1-{\epsilon}$. Thus the transformation (\ref{ac}) relates equation (\ref{ae}) to another Heun equation if ${{q}\over{{\alpha}{\beta}}}$ is equal to $0,1,d$. Note that in this case the transformation (\ref{ac}) 

$$({{dy}\over{dx}}{{1}\over{y}})({{du}\over{dx}}{{1}\over{u}})=f(x)$$

involves solutions to a Heun type equation and their first derivative only. No other formula is known relating a solution to a Heun equation to its first derivative. For hypergeometric functions this type of relation is known : the first derivative of a hypergeometric function is proportional to another hypergeometric function, see section 3. \\
\\

Now using the transformation (\ref{ad}) will try to find another relation between equation (\ref{ae}) and a Heun type equation.
In order to achieve this we will use equations (\ref{aca}) and (\ref{acb})  
together with the relation (\ref{ad}).

Let us perform the following change of variable in equation (\ref{aca})

$$x={\int{{\eta}(\xi){d{\xi}}}}$$

in order to obtain

\begin{equation}
\label{af}
{{dh_1}\over{d{\xi}}}-({{{\eta}{\alpha}}\over{f_1}}){{h_1}^2}+[{\eta}(\xi){F_1}(\xi)-{{df_1}\over{d{\xi}}}{{1}\over{f_1}}]{h_1}-{\eta}(\xi){f_1}(\xi)=0
\end{equation}

with ${h_1}(\xi)=h({\int{{\eta}(\xi){d{\xi}}}})$, ${f_1}(\xi)=f({\int{{\eta}(\xi){d{\xi}}}})$ and ${F_1}(\xi)=F({\int{{\eta}(\xi){d{\xi}}}})$.

Now let

$${\eta}(\xi){F_1}(\xi)=g(\xi)$$

$${\alpha}(\xi)=-{{f_1}\over{\eta}}$$

and suppose the following relation holds

$$ {{\eta}{f_1}}={{\eta}(\xi)}f({\int{{\eta}(\xi){d{\xi}}}})=H(\xi)$$

The last equation is a differential equation for ${\eta}(\xi)$, that can be solved as follows. Let 

\begin{equation}
\label{ag}
{\eta}(\xi)={{du}\over{d{\xi}}}
\end{equation}
such that the above equation becomes 

$${{du}\over{d{\xi}}}f(u)=H(\xi)$$

which can be solved, given the function $f(u)$, by resolving 

\begin{equation}
\label{aga}
{\int{f(u){du}}}={\int{{H}(\xi){d{\xi}}}}
\end{equation}
w.r.t. $u(\xi)$. From equation (\ref{ag}) one has the following relation between the variable $x$ and $\xi$ :

\begin{equation}
\label{ah}
x={u(\xi)}
\end{equation}

The previous relations imply that equation (\ref{af}) becomes 

\begin{equation}
\label{ai}
{{dh_1}\over{d{\xi}}}+{{h_1}^2}+[g(\xi)-{{df_1}\over{d{\xi}}}{{1}\over{f_1}}]{h_1}-H(\xi)=0
\end{equation}

We further let 

\begin{equation}
\label{aj}
{f_1}(\xi)={e^{\int{(g-G){d{\xi}}}}}
\end{equation}

such that equation (/ref{ai}) for $h_1$ becomes

$${{dh_1}\over{d{\xi}}}+{{h_1}^2}+G(\xi){h_1}-H(\xi)=0$$

If we define ${h_1}(\xi)={{dk}\over{d{\xi}}}{{1}\over{k}}$, then the above equation becomes

\begin{equation}
\label{ak}
{{{d^2}k}\over{d{\xi}^2}}+G(\xi){{dk}\over{d{\xi}}}-H(\xi)k=0
\end{equation}

We would like this equation to be a Heun equation. To achieve this we 
let
$$G(\xi)={{1}\over{\xi}}{{a{c_1}({\xi}-{\lambda})+{\xi}(b-a{\mu})}\over{m{c_1}({\xi}-{\lambda})+{\xi}(n-m{\mu})}}-{{c}\over{\xi}}$$
and 

$$H(\xi)={{D}\over{{\xi}({\xi}-{\lambda})}}$$

with $a$, $b$, $c$, $c_1$, $D$, $m$, $n$, $\mu$ and $\lambda$ being constants.
It is easy to see that equation (\ref{ak}) can be reduced to a Heun equation of form (\ref{ae}), by a change of variable ${\xi} \rightarrow {\lambda}{\xi}$. In this case the regular singularities are located at $0, 1, d={{m{c_1}}\over{m{c_1}-m{\mu}+n}}$ and ${\infty}$. The other parameters are given by 

$${\gamma}={{a-mc}\over{m}}$$
$${\delta}=0$$
$${\epsilon}=[a{c_1}+b-a{\mu}-cm{c_1}-c(n-m{\mu})-({{a-mc}\over{m}})(m{c_1}+n-m{\mu})]{{1}\over{(m{c_1}+n-m{\mu})}}$$
$${\alpha}{\beta}=-D$$
$$q={{-Dm{c_1}}\over{m{c_1}-m{\mu}+n}}=-Dd$$

Now, choose the function $f(x)$ to be 
$f(x)={{D/{\lambda}}\over{x+{\mu}}}$
which by equation (\ref{ah}) gives 
\begin{equation}
\label{al}
f(u)={{D/{\lambda}}\over{u+{\mu}}}
\end{equation}

such that we can solve equation (\ref{aga}). This will give 

$${\ln}(u+{\mu})={\ln}({{{\xi}-{\lambda}}\over{\xi}})+{c'}$$
where ${c'}$ is an arbitrary integration constant.
This means we have 
\begin{equation}
\label{am}
u={c_1}{{{\xi}-{\lambda}}\over{\xi}}-{\mu}
\end{equation}
with ${c_1}={e^{c'}}$
Using (\ref{ah}) and (\ref{am}) we have the following relation between $x$ and $\xi$,

\begin{equation}
\label{an}
{\xi}={{\lambda}\over{1-{c_2}(x+{\mu})}}
\end{equation}

with ${c_2}={{1}\over{c_1}}$. Equations (\ref{ag}) and (\ref{am}) allows us to find ${\eta}(\xi)$,

\begin{equation}
\label{ao}
{\eta}(\xi)={{du}\over{d{\xi}}}={{{c_1}{\lambda}}\over{{\xi}^2}}
\end{equation}
We also have the function $f(x)={f_1}(\xi)$ using (\ref{al}) and (\ref{am}), 

$${f_1}({\xi})={{(D/{c_1}{\lambda}){\xi}}\over{{\xi}-{\lambda}}}$$

so that one can write equation (\ref{acb}) : to achieve this one needs to calculate ${F_1}({\xi})={{g(\xi)}\over{{\eta}({\xi})}}$ and ${\alpha}(\xi)=-{{{f_1}(\xi)}\over{{\eta}(\xi)}}$ as functions of the variable $x$. Using (\ref{aj}) and (\ref{an}) we have  
\begin{equation}
\label{ap}
{({{g(\xi)}\over{{\eta}({\xi})}})_{x}}={{N(x)}\over{[1-{c_2}(x+{\mu})](mx+n)(x+{\mu})}}
\end{equation}
with 
$$N(x)={x^2}({{a}\over{c_1}}-m{{c-1}\over{c_1}})+{x}({{b}\over{c_1}}+{{{\mu}a}\over{c_1}}-n{{c-1}\over{c_1}}-m)+{{b{\mu}}\over{c_1}}-{{c-1}\over{c_1}}n{\mu}-n$$

\begin{equation}
\label{aq}
{({{{f_1}(\xi)}\over{{\eta}(\xi)}})_{x}}={{D}\over{c_1}}{{1}\over{{[1-{c_2}(x+{\mu})]^2}(x+{\mu})}}
\end{equation}
where ${(.)_{x}}$ denotes evaluation in function of the variable $x$ using the relation (\ref {an}), such that equation (\ref{acb}) becomes 

\begin{equation}
\label{aqa}
{{{d^2}y}\over{dx^2}}-{({{g(\xi)}\over{{\eta}({\xi})}})_{x}}{{dy}\over{dx}}- {({{{f_1}(\xi)}\over{{\eta}(\xi)}})_{x}}y=0
\end{equation}

which has four regular singular points located at $-{\mu}$, $-{{n}\over{m}}$, 
${{1-{c_2}{\mu}}\over{c_2}}$ and $\infty$.

Thus equation (\ref{aqa}) is related to equation (\ref{ak})
$${{{d^2}k}\over{d{\xi}^2}}+G(\xi){{dk}\over{d{\xi}}}-H(\xi)k=0$$
via the relation (\ref{ad}) where all functions are expressed as functions of the variable $x$, 
\begin{equation}
\label{ar}
({{dy}\over{dx}}{{1}\over{y}})h(x)=f(x)
\end{equation}

with $h(x)=({{dk}\over{d{\xi}}}{{1}\over{k}})_{x}$

This relation also involves solutions to a Heun type equation and their first derivative, as is the case for the transformation (\ref{ac}) mentioned earlier in this section.

\section{A new Heun-to-Hypergeometric transformation}

Equation (\ref{aqa}) reduces to an equation with {\it three} regular singularities when for instance 
\begin{equation}
\label{as}
N(x)={R(mx+n)^2}
\end{equation}

with $R$ being a constant. In this case the transformation from equation (\ref{ak}) to equation (\ref{aqa})
is a transformation between a Heun equation to a hypergeometric type equation, with 

$${({{g(\xi)}\over{{\eta}({\xi})}})_{x}}={{R(mx+n)}\over{[1-{c_2}(x+{\mu})](x+{\mu})}}$$

$$ {({{{f_1}(\xi)}\over{{\eta}(\xi)}})_{x}}={{D}\over{c_1}}{{1}\over{{[1-{c_2}(x+{\mu})]^2}(x+{\mu})}}$$

such that equation (21) now has has three regular singularities located at $x=-{\mu}$, ${{1-{c_2}{\mu}}\over{c_2}}$ and $\infty$.
  
In order to satisfy equation (\ref{as}) one will solve the following system, obtained using the expressions in equation (\ref{ap})

\begin{equation}
\label{asa}
{{a}\over{c_1}}-{{c-1}\over{c_1}}m=R{m^2}
\end{equation}

\begin{equation}
\label{asb}
{{b+{\mu}a}\over{c_1}}-{{c-1}\over{c_1}}(m{\mu}+n)-m=2mnR
\end{equation}

\begin{equation}
\label{asc}
{{b{\mu}}\over{c_1}}-{{n{\mu}(c-1)}\over{c_1}}-n={n^2}R
\end{equation}

Equation (\ref{asa}) gives for $c_1$

$${c_1}={{a-(c-1)m}\over{R{m^2}}}$$

Equation (\ref{asb}) gives for $\mu$

$${\mu}={{(a-(c-1)m)(2mnR+m)+(-b+(c-1)n)R{m^2}}\over{(a-(c-1)m)R{m^2}}}$$

Equation (\ref{asc}) is a linear equation for $R$ which gives 

$$R={{{L^2}n-(am-l{m^2})K}\over{(2amn-b{m^2}-l{m^2}n)K-{L^2}{n^2}}}$$

with $l=c-1$, $L=a-lm$, $K=b-nl$.

Note that although the Heun equation (\ref{ak}) has ${\alpha}{\beta}=-D$ and $q={{-Dm{c_1}}\over{m{c_1}-m{\mu}+n}}=-Dd$
, the parameter ${\epsilon} \neq 0$. If one would have ${\alpha}{\beta}=-D$ and $q={{-Dm{c_1}}\over{m{c_1}-m{\mu}+n}}=-Dd$
and ${\epsilon}=0$ then the Heun equation (\ref{ak}) would 'loose' the singular point at $x=d$ and would be equivalent to a hypergeometric equation. Our calculation thus shows that one can use the non-linear transformation (\ref{ad}) to relate a Heun equation (\ref{ak}) with ${\alpha}{\beta}=-D$, $q={{-Dm{c_1}}\over{m{c_1}-m{\mu}+n}}=-Dd$, ${\epsilon} \neq 0$, ${\gamma} \neq 0$, ${\delta}=0$, having four regular singularities (see previous section), to a hypergeometric type equation having three regular singularities.

\section{Acknowledgements}

We thank Prof. R. Maier for his useful comments on this paper and we are very grateful to Prof. A. Noels for allowing the use of a computer at the Institut d'Astrophysique et de G\'{e}ophysique at the Universit\'{e} of Liege (Belgium).

\end{document}